\documentclass[aps,prb,showpacs,twocolumn]{revtex4}
\usepackage{graphicx}
\begin{document}

\newcommand{\eqn}[1]{(\ref{#1})}

\newcommand{\be}{\begin{equation}}
\newcommand{\ee}{\end{equation}}
\newcommand{\bea}{\begin{eqnarray}}
\newcommand{\eea}{\end{eqnarray}}
\newcommand{\bean}{\begin{eqnarray*}}
\newcommand{\eean}{\end{eqnarray*}}

\newcommand{\nn}{\nonumber}




\title{Spin Hall Effect and Spin Orbit coupling in Ballistic Nanojunctions }
\author{S. Bellucci $^a$ and P. Onorato $^a$ $^b$ \\}
\address{
        $^a$INFN, Laboratori Nazionali di Frascati,
        P.O. Box 13, 00044 Frascati, Italy. \\
        $^b$Department of Physics "A.Volta", University of Pavia, Via Bassi 6, I-27100 Pavia, Italy.}
\date{\today}
\begin{abstract}
We propose a new scheme of spin filtering based on nanometric
crossjunctions in the presence of Spin Orbit interaction, employing
ballistic nanojunctions patterned in a two-dimensional electron gas.
We demonstrate that the flow of a longitudinal unpolarized current
through a ballistic X junction patterned in a two-dimensional
electron gas with Spin Orbit coupling (SOC) induces a spin
accumulation which has opposite signs for the two lateral probes.
This spin accumulation, corresponding to a transverse pure spin
current flowing in the junction, is the main observable signature of
the spin Hall effect in such nanostructures.

We benchmark the effects of two different kinds of Spin Orbit
interactions. The first one ($\alpha$-SOC) is due to the interface
electric field that confines electrons to a two-dimensional layer,
whereas the second one ($\beta$-SOC) corresponds to the interaction
generated by a lateral confining potential.

\end{abstract}


\maketitle

{\it Introduction.} The classical Hall effect occurs when an
electric current flows through a conductor subjected to a
perpendicular magnetic field. In this case   the Lorentz force
deflects the electrons, and charge builds up on one side of the
conductor, resulting in an observable Hall voltage\cite{Hall}.

In the absence of an external magnetic field, some  unconventional
Hall-type effects involving the electron spin become possible in
 systems with Spin Orbit (SO) interactions, such as the
 spin Hall effect (SHE),   predicted by theorists over 30 years
ago\cite{[3],Hir}. In analogy with the conventional Hall effect, an
external electric field can be expected to induce a pure
transverse spin current, in the absence of applied magnetic fields.
\begin{figure}
\includegraphics*[width=.95\linewidth]{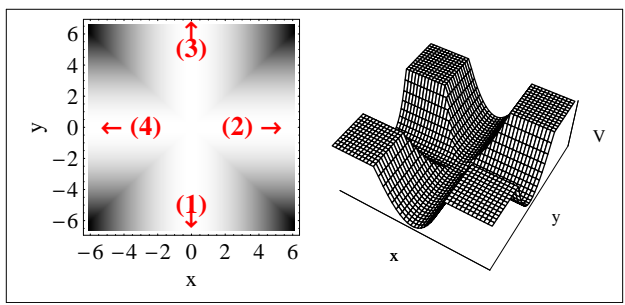}
 \caption { Density  and 3D Plots of the potential  $V_{c}(x,y)$ which describes the nanojunction.
 The junctions, in analogy with the  devices proposed in ref.[\onlinecite{Nprl}],  can be assumed
 as  crossing junctions between two Q1D wires of width $W$ ranging between  $\sim
25$  nm  and $100$ nm. }
\end{figure}
In fact, the opposite spins can be separated and then accumulated on
the lateral edges (probes 2 and 4 of Fig.(1)), when they are
transported by a pure spin Hall current flowing in the transverse
direction, in response to an unpolarized charge current in the
longitudinal direction (injected in probe 1 of Fig.(1)).

When the scattering due to the impurities is spin-dependent, spin-up
and spin-down  electrons of an unpolarized beam are scattered into
opposite directions, resulting in spin-up and spin-down charge Hall
currents.  The presence of the  accumulation of spins shows that the
SHE exists, but in this case it is {\em extrinsic} because it
originates from spin-dependent scattering\cite{[3]}. It was realized
long ago that the extrinsic spin Hall current is the sum of two
contributions\cite{Nozieres}. The first contribution (commonly known
as ``skew-scattering" mechanism \cite{Smit55,Smit58}) arises from
the asymmetry of the electron-impurity scattering in the presence of
spin-orbit interactions \cite{Mott},  the second one, i.e. the
so-called ``side-jump" mechanism
\cite{Berger70a,Berger70b,Lyo72,Nozieres}, is caused by the
anomalous  relationship between the physical and the canonical
position operator\cite{vignale}.

 More recently, it has been pointed out that
there may exist a different, purely {\it intrinsic SHE}. Recent
theoretical arguments have unearthed the possibility for pure
transverse spin Hall current, that is several orders of magnitude
larger than in the case of the extrinsic effect, arising due to
intrinsic mechanisms, related to the spin-split band structure in
SO coupled bulk \cite{[6],[7]} or mesoscopic \cite{68}
semiconductor systems. Often the spin currents are not directly
observable, and so their detection requires measuring the spin
accumulation deposited by the spin currents at the sample
edges\cite{29}.

\

Thus, the SO coupling  plays a central role in the SHE phenomenology
and its properties were largely investigated in two-dimensional
electron systems.  Analogously to the case of the Hall effect, the
SHE is based on a
 velocity dependent force, such as the one given by the SO interaction due to
 the presence of an external electric field, which,
 unlike the magnetic field,  does not break
the time reversal symmetry.
 The SO Hamiltonian due to an electric field, ${\bf E}({\bf r})$, is
given by \cite{Thankappan}
\begin{equation}
\hat H_{SO} = -\frac{\lambda_0^2}{\hbar}\;m_0e{\bf E}({\bf r})\cdot
\left[\hat{{\bf \sigma}}\times \hat{\bf p}\right]. \label{H_SO}
\end{equation}
Here  $m_0$ is  the electron mass in vacuum,
 $\hat{{\sigma}}$ are
the Pauli matrices, {$\hat{\bf p}$ is the canonical momentum
operator ${\bf r}$ is a 3D position vector } and $\lambda_0^2=
\hbar^2/(2m_0 c)^2$.

In the present work we consider low dimensional electron systems
formed by quasi-one-dimensional (Q1D) devices patterned in 2DEGs
entrapped in a  potential well at the interface of a
heterostructure. Thus, $m_0$ and $\lambda_0$ are substituted by the
effective values $m^*$ and $\lambda$ they take in the material. In
2DEGs
 there are  different types of SO
interaction\cite{morozb}, such as the  Dresselhaus term which
originates from the inversion asymmetry of the zinc-blende
structure\cite{3t},  the Rashba ($\alpha$-coupling) term due to the
quantum-well potential~\cite{Kelly,Rashba,Bychkov} that confines
electrons in the 2DEG, and  the confining ($\beta$-coupling) term
arising from the in-plane electric potential that is applied to
squeeze the 2DEG into a quasi-one-dimensional
channel~\cite{Thornton,Kelly}.

A useful way to describe the effect of the SOC is to use an
effective magnetic field,  coplanar to the motion plane   and
orthogonal to the electrons speed for the $\alpha$ term, and
perpendicular to the motion plane for the $\beta$ term.

 \

In finite-size systems with SO couplings, there is always the
possibility that spin Hall phenomenology (accumulation of opposite
sign on the lateral edges of 2-terminal devices, or transverse
spin currents in 4-terminal devices) is generated by some kind of
edge effect. For the Rashba  SO coupled systems this has been
discussed recently in ref.\cite{ref1}, where spin-charge coupled
transport was studied in disordered systems. It follows that the
spin accumulation generally depends on the nature of the boundary
and therefore the SHE in a SO coupled system can be viewed as a
non-universal edge phenomenon. Analogously, the spin polarization
induced by a current flow in a 2DEG was assumed as a geometric
effect originating from special properties of the electron
scattering at the edges of the sample\cite{ref2}. Moreover, in
some recent papers\cite{noish,qse,hatt}  it has been argued  that
the confining potential induced SO coupling will generate spin
Hall like phenomenology, and it was suggested that the $\beta$
coupling yields stronger effects.

\

In order to discuss the different effects of the two SO terms, we
formulate a B\"uttiker-Landauer approach\cite{BL}
 to the spin accumulation problem in a four terminal ballistic nanojunction,
 by studying the electron transport at quantum wire (QW)
 junctions in the strongly coupled regime\cite{kirc}.

 The four-probe cross junction system appears to be like an
ultra-sensitive scale, capable of reacting to the smallest
variations of the external\cite{kirc} or effective magnetic
field\cite{noiq,Nprb}. In fact, any breakdown of the symmetry
(left right $12-14$ in Fig.(1)) produces a transverse current
(charge Hall current for the external magnetic field and pure spin
current for the effective field due to the SOC). Since the
effective fields due to the SOC are usually small, such devices
can be quite useful, in order to obtain a spin polarized current,
and have been extensively studied in recent
years\cite{Nprl,noiq,Nprb,Nprin}.

 \

{\it SO interaction in  quasi-one-dimensional systems} -
 The ballistic one-dimensional wire is a nanometric  solid-state device in
 which the transverse motion (along $x$) is quantized into discrete modes,
 and the longitudinal   motion ($y$ direction) is free.
In this case, electrons propagate freely down to a clean narrow
pipe and electronic transport with no scattering can occur.

In line with the refs.\cite{3840}, the lateral confining potential
of a QW,$V_c(x)$, is approximated by a parabola \bea
\hat{H}_0=\frac{{\bf p}^2}{2 m^*}+V_c({\bf r})=\frac{{\bf p}^2}{2
m^*}+\frac{m^*}{2}\omega^2 x^2. \label{h0} \eea  The quantity
$\omega$ controls the strength (curvature) of the confining
potential while  the in-plane electric field $e{\bf E}_{c}({\bf
r})=-\nabla V_{c}({\bf r})=-m^*\omega^2{x}$ is  directed along the
transverse direction.

We assume that the SO interaction Hamiltonian $\hat{H}_{SO}$ in
eq.~(\ref{H_SO}) is formed by two contributions
$\hat{H}_{SO}=\hat{H}_{SO}^\alpha+\hat{H}_{SO}^\beta$. The first
one,
\begin{equation}
\hat{H}_{SO}^\alpha = \frac{\alpha}{\hbar}
\left(\hat{{\sigma}}\times\hat{\bf p}\right)_z=
i\alpha\left(\sigma_y\frac{\partial}{\partial x}
-\sigma_x\frac{\partial}{\partial y}\right), \label{H_a}
\end{equation}
arises from  the interface-induced (Rashba) electric field that can
be reasonably assumed to be uniform and directed along the $z$-axis.
The SO-coupling constant $\alpha$ takes values within the range
$\sim 10^{-11}-10^{-12}$ eV m, for different
systems~\cite{Das,Nitta,Luo,Bychkov,Hassenkam}.

\

The second contribution $\hat{H}_{SO}^\beta$ to $\hat H_{SO}$ comes from
the parabolic confining potential
\begin{equation}
\hat{H}_{SO}^\beta = \frac{\beta}{\hbar}\frac{x}{l_\omega}
\left(\hat{{\sigma}}\times \hat{\bf p}\right)_x =
i\beta\frac{x}{l_\omega}\sigma_z \frac{\partial}{\partial y}.
\label{H_b}
\end{equation}
Here $l_\omega=(\hbar/m^*\omega)^{1/2}$ is the typical spatial scale
associated with the potential $V_{c}$ and $\beta\equiv\lambda^2
{m^*}\omega^2 l_\omega$. A comparison of typical electric fields
originated from the quantum-well and lateral confining potentials
allows one to conclude that a plausible estimate for $\beta$ should
be roughly $\sim 0.1 \alpha$[\onlinecite{morozb}] (in a InGaAs based
heterostructure with QWs of width $\sim 100 nm$). Moreover, in
square quantum wells where the value of $\alpha$ is considerably
diminished~\cite{Hassenkam,Chen} (by one order of magnitude) the
constant $\beta$ may well compete in size with $\alpha$.

\

{\it Eigenfunctions in a QW with $\alpha$ interaction }- As we
know from ref.\cite{me}, the QW Hamiltonian in the presence of
$\alpha$-SOC  cannot be  exactly diagonalized. We can calculate
its spectrum (and related wavefunctions) by a numerical
calculation or with simple perturbation theory. In order to get to
our result, we analyze the Hamiltonian eq.(\ref{H_a}) in the
general case and separate the commuting part
($[\hat{H}_c,\hat{H}_0]=0$),
\begin{equation}\label{hd}
  \hat{H}_c=\frac{\alpha}{\hbar}p_y
  \hat{\sigma}_x=\frac{p_R p_y}{ m^*}\hat{\sigma}_x,
\end{equation}
where $p_R=\hbar k_R\equiv \frac{m^* \alpha}{\hbar}$.   The  non diagonal term,
\begin{equation}\label{hnd}
  \hat{H}_n=\frac{\alpha}{\hbar} p_x
  \hat{\sigma}_y= \frac{\hbar p_R}{m^* l_\omega}
(\hat{a}_x-\hat{a}^\dag_x)(\hat{\sigma}^x_++\hat{\sigma}^x_-),
\end{equation}
where $\hat{a}_x,\hat{a}^\dag_x$ are the creation-annihilation
operators of the 1D quantum mechanical harmonic
oscillator\cite{mess}, can be neglected in the first order
approximation and becomes relevant just near the crossing points
$\pm k_c$\cite{notakc}, as  it was discussed in
refs.[\onlinecite{me, governale }] and bibliography therein.

It follows that the Rashba subbands splitting in the energies, in
the first order approximation, read
\begin{equation}\label{er}
 \varepsilon_{n,k,s_x} =\hbar
\omega(n+\frac{1}{2})+\frac{\hbar^2}{2m^*}((k\pm k_R)^2- k_R^2).
\end{equation}
Here $s_x=+ 1$($-1$) corresponds to $\chi_\rightarrow$
($\chi_\leftarrow$)  spin eigenfunctions along the $x$ direction.
Hence
 we can conclude that 4-split
channels are present for a fixed Fermi energy, $\varepsilon_F$, corresponding to $\pm
p_y$ and $s_x=\pm1$ with eigenfunctions
$$
\varphi_{\varepsilon_F,n,s_x}=u_n(x)e^{i k y}\chi_{s_x},
$$
where $u_n(x)$ are displaced harmonic oscillator eigenfunctions.

\

{\it Eigenfunctions in a QW with $\beta$ interaction }- As it was
discussed  in the  case of a QW, the effect of the $\beta$-SOC is
analogous to the one of a uniform effective magnetic field,
\begin{equation} {B}_{eff} =
\frac{\lambda^2}{\hbar}\;\frac{{m^*}^2\omega^2
c}{e}\equiv\frac{\beta}{\hbar l_\omega}\frac{m^*c}{e},\label{Beff}
\end{equation} orthogonal to the 2DEG directed upward or
downward, according to the spin polarization along the $z$
direction.

Next we introduce  $\omega_{\text{eff}}=\frac{\beta}{\hbar
l_\omega}$, $\omega_0^2=\omega^2-\omega_{\text{eff}}^2$ and the
total frequency $\omega_T=\sqrt{\omega^2+\omega_{\text{eff}}^2}$,
thus
\begin{equation}\label{hnws3}
\hat{H}_0+\hat{H}^\beta_{SO} =
\frac{\omega_0^2}{\omega_T^2}\frac{p_y^2}{2m^*}+\frac{p_x^2}{2m^*
}+\frac{m^* \omega_T^2}{2}(x-x_0)^2,
\end{equation}
where $x_0=s \frac{\omega_c\omega_{\text{eff}} p_y}{\omega_T^2
m^*}$, $s=\pm1$, corresponds to the spin polarization along the
$z$ direction. 4 spin channels are present for a fixed Fermi
energy $\varepsilon_F$, corresponding to $\pm p_y$ and $s_z=\pm1$
with wavefunctions
$$
\varphi_{\beta,\varepsilon_F,n,s_z}(x,y)=u_n(x+s_z x_0)e^{i k y}\chi_{s_z}.
$$

{\it The junction }- Starting from the model of a QW, we are able to
write the model potential energy of the electrons in the x-y plane
as $V_c(x,y)=\frac{m^* \omega^2}{2}x^2$ for $|y|>|x|$ and
$V_c(x,y)=\frac{m^* \omega^2}{2}y^2$ for $|x|>|y|$, for QWs running
along the x and y axes as in Fig.(1) (see ref.[\onlinecite{kirc}]).

Next we follow the quantum mechanical approach to the calculation
of electron scattering proposed by Kirczenow in
ref.[\onlinecite{kirc}]. This approach is based on the analytic
solution of the quantum-mechanical Schr\"odinger equation in each
of the four QWs.

 \

{\it $\alpha$-Coupling }- In each of the four QWs, $H^\alpha$  has
eigenfunctions belonging to 1D subband ($n=0,1,...$) given by \bea
f^\parallel_{n,q,\rightarrow}=u_n(x)e^{i q y}e^{i k_R y}\chi^x_\rightarrow \quad f^\parallel_{n,q,\leftarrow}=u_n(x)e^{i q y}e^{-i k_R y}\chi^x_\leftarrow ,\nn\\
f^=_{n,q,\rightarrow}=u_n(y)e^{i q x}e^{i k_R x}\chi^y_\rightarrow
\quad f^=_{n,q,\leftarrow}=u_n(y)e^{i q x}e^{-i k_R
x}\chi^y_\leftarrow  \nn, \eea where $\parallel$ stands for wire 1
or 3 and $=$ for wire 2 or 4. An electron in subband $n$  having
spin $s_x=s$ and  wavevector
$$
k_w^s(\varepsilon_F,n)=s k_R\pm \sqrt{k_R^2+k_n^2} \equiv s k_R \pm
q,
$$
where  $ k_n^2=2 m^*\left(\varepsilon_F-\hbar
\omega_T(n+\frac{1}{2})\right)$  {\bf and $q=\sqrt{k_R^2+k_n^2}$},
is incident on the junction from wire 1.  The electron eigenstate
is given by $\psi=\psi^i$ in wire $i=1,2,3,4$ respectively, where
$\psi^1=f^1_{n,k,s}+\sum_{r,s'} a^1_{r,s'} f^1_{r,-k,s'}$ and
$\psi^j=\sum_r a^j_{r,s'} f^j_{r,\pm k,s'}$ for $j=2,3,4$ ($+$ for
$j=2$ and $3$, $-$ for $j=4$). The sums are over all subbands,
including those with imaginary $q$ (evanescent partial waves). The
expansion coefficients $a^i_{r,s'}$, and the scattering
probabilities $T_{i,j}^{s,s'}$, were found numerically from the
continuity of $\psi$ and $\nabla \psi$ at $x=\pm y$.

If we limit ourselves to the lowest subband ($n=0$),  the out of
plane $\langle S_z ({\bf r})\rangle$ component of the spin
accumulation (local spin density) can be obtained in each probe as
a simple function of the expansion coefficients: \bea \label{dd}
\langle S^i_z (\xi, \eta)\rangle=\frac{\hbar}{2}S^i_0 \cos(2 k_R
\eta+\delta_i) .\eea Here
$S^i_0=|a^i_{0,\rightarrow}(a^i_{0,\leftarrow})^*|$, $\delta_i=
Arg\left(a^i_{0,\rightarrow}(a^i_{0,\leftarrow})^*\right)$ while
$(\xi, \eta)$ are $(y,x)$ in probe 2, $(x,y)$ in probe 3 and
$(y,-x)$ in probe 4. From eq.(\ref{dd}) we obtain the known spin
oscillations, upon which the Datta-Das device for the spin
filtering is also based\cite{me}. These oscillations can disappear
in a device like the one proposed in ref.\cite{Nprl}, where the SO
interaction vanishes out of the crossing zone. When a spin
unpolarized current is injected in lead 1, it yields $S^2_0=S^4_0$
and $S^3_0=0$, as a consequence of the symmetry of the system.

Next let us turn to considering the presence of opposite spin
polarizations in the two opposite probes 2 and 4 of the
nanojunction. The phase difference $\delta_2 -\delta_4$ is
responsible for a net spin polarization of the current transmitted
in the two transverse leads. If we introduce two collectors at a
distance $L_c< \frac{\pi}{4 k_R}$ from the center of the junction,
no oscillations are present.

\begin{figure}
\includegraphics*[width=.95\linewidth]{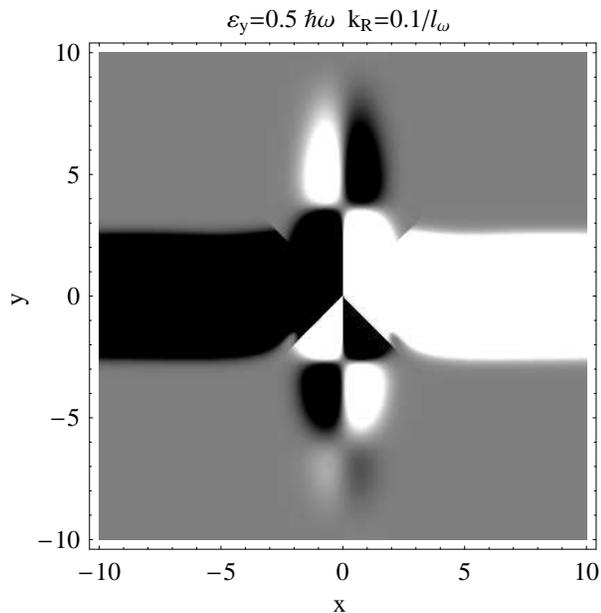}
 \caption { The
out-of-plane component $\langle S_z({\bf r})\rangle$ of the
 spin accumulation induced by the quantum
transport of the unpolarized charge current injected from the lead 1
into a 4-terminal cross junction ($k_R l_\omega=0.1$). This picture
shows how lateral spin-$\uparrow$ (white) and spin-$\downarrow$
(black) densities flow in opposite directions through the transverse
 leads to generate a linear response spin Hall
current $I_{sH}$, which changes sign upon reversing the bias
voltage (i.e. injecting the current in lead 4). Here we display
only the current exiting from the junction, thus we observe
unpolarized currents in leads 1 and 3. Since the local spin
density is due to the evanescent states, it is negligeable
everywhere, except in proximity of the center of the junction. }
\end{figure}
Thus, as shown in Fig.(2), obtained taking into account the first
5 subbands, lateral spin-$\uparrow$ and spin-$\downarrow$
densities will flow in opposite directions through the  transverse
leads. This behaviour is unchanged if we substitute the collectors
by ideal leads with vanishing SO interaction.

For evaluating the order of magnitude corresponding to the spin
polarization, we can calculate the dependence of $S_z$ on the
strength of the Rashba coupling.
\begin{figure}
\includegraphics*[width=.5\linewidth]{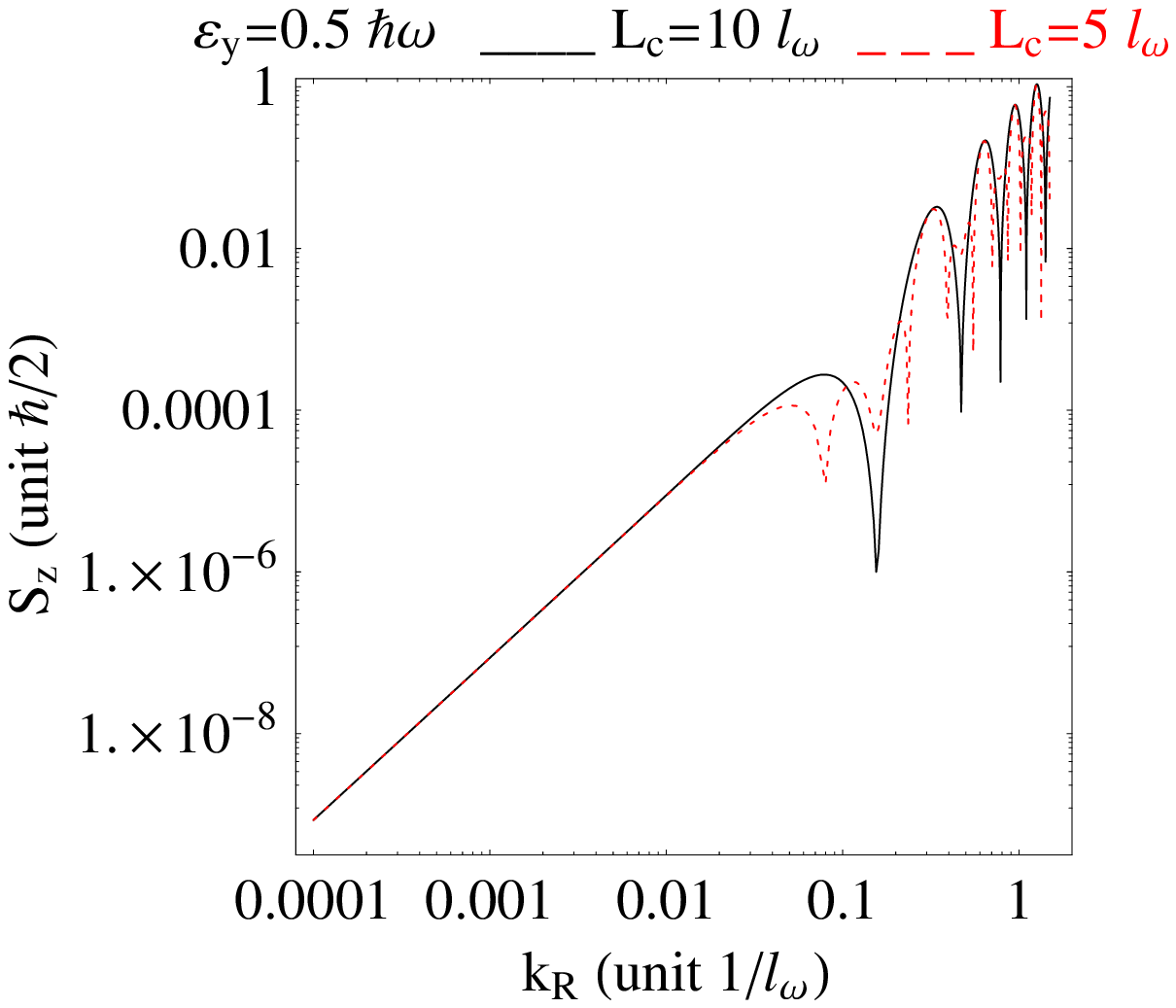}
\includegraphics*[width=.435\linewidth]{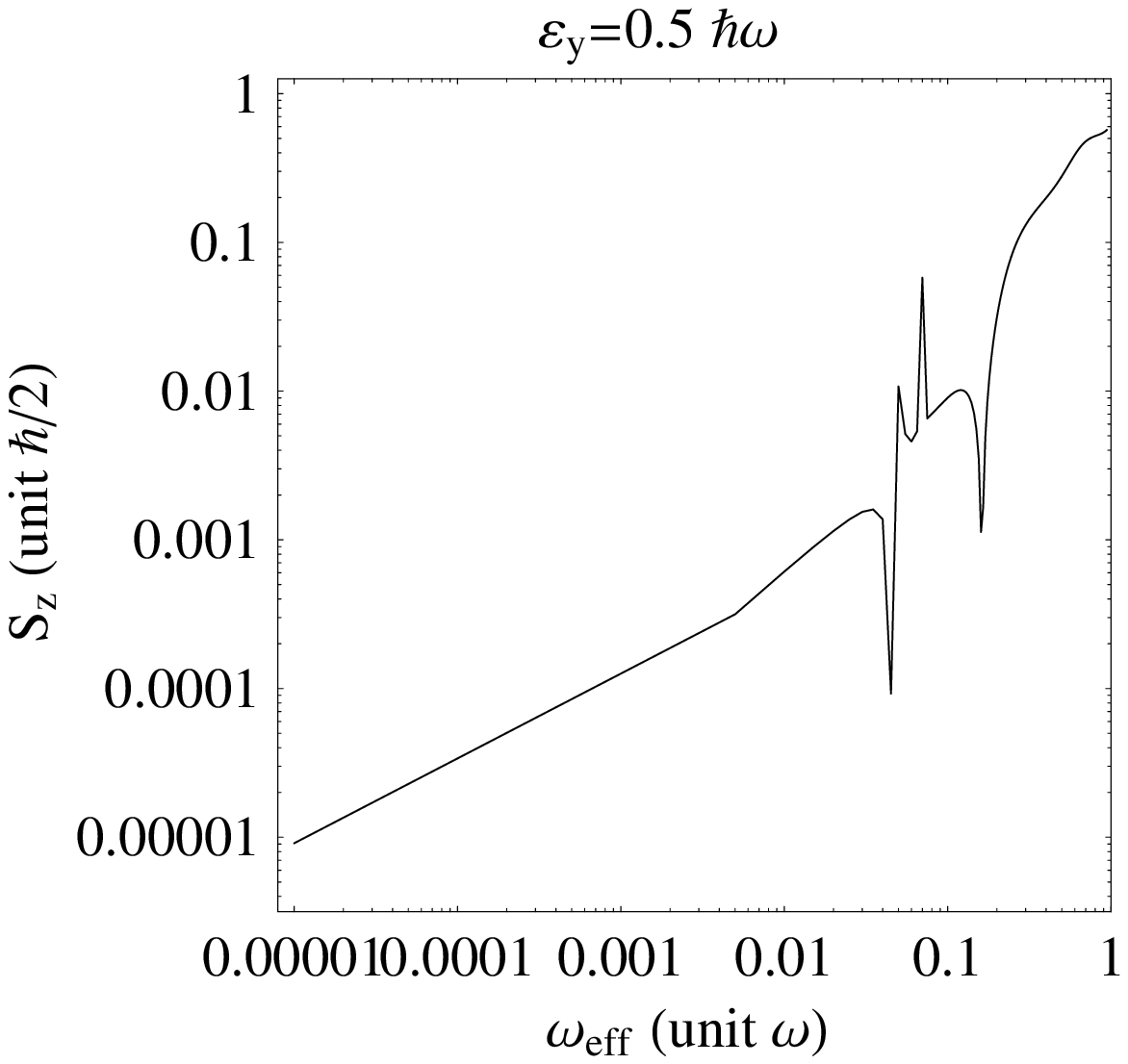}
 \caption {The dependence of the maximum value of the out-of-plane component of
 the spin accumulation on the strength of the SO coupling (reported in logarithmic scale).
 On the left we display the case of $\alpha$-SOC
 where $S_z$ is reported as a function of $k_R$. On the right we show the case of $\beta$-SOC
 where the coupling is expressed in terms of $\omega_{\text{eff}}$ in units of ($\omega$).
 Notice that the abscissa of the two graphs is the same, in fact $k_R
l_\omega=\frac{\alpha}{l_\omega \hbar \omega}$ while
$\frac{\omega_{eff}}{\omega}=\frac{\beta}{ l_\omega \hbar \omega
}$. }
\end{figure}
In the left panel of Fig.(3) we show the spin polarization, as a
function of the Rashba wavevector $k_R$, for two different values
of the distance between the collectors and the center of the
junction, i.e. $L_c=5l_\omega$ and $L_c=10 l_\omega$.

\

Next we discuss some realistic or theoretical devices, capable of
acting as spin filters based on the Rashba interaction. In  usual
devices\cite{morozb}, the natural value reads $\alpha_0\sim
10^{-11} eV\, m$ and we obtain $k_R\sim 10^{-4} nm^{-1}$, whereas
a typical nanojunction should be made by using narrow QWs of a
width, W, from $25$ to $75 nm$, corresponding to $l_\omega\sim 5$
to $15 nm$. It follows that $k_R l_\omega$ ranges between $
10^{-3}$ and $ 10^{-2}$, corresponding to a spin value $\langle
S_z\rangle$ between $10^{-7}$ and $10^{-5}$ $\hbar/2$, as we show
in Fig.(3).

In the device proposed in ref.[\onlinecite{Nprl}] the value of
$\alpha$ can take values up to  $\sim 6\times 10^2 \alpha_0$, while
$W$ is $\sim 25 nm$, and we have the value $k_R l_\omega \lesssim
0.3$, corresponding to a spin value $\langle S_z\rangle$ which
ranges from $10^{-4}$ to $10^{-3}$ $\hbar/2$, in agreement with the
results reported there.

\

{\it $\beta$-Coupling }- In each of the four QWs $H^\beta$  has
eigenfunctions belonging to a 1D subband ($n=0,1,...$) given by \bea
f^\parallel_{n,q,s}=u_n(x-s x_0)e^{i k y}\chi^z_\uparrow \nn\\
f^=_{n,q,s}=u_n(y-s x_0)e^{i k x}\chi^z_\downarrow \nn. \eea An
electron in subband $n$, having spin $s_z=s$ and  wavevector $
k(\varepsilon_F,n)$, is incident on the junction from wire 1. We
can evaluate the transmission coefficient from the expansion
coefficients as we did above, whereas in this case no oscillations
will be observed in the spin polarization ($[H^\beta,S_z]=0$).

Here we limit ourselves to the lowest subband ($n=0$), thus the out
of plane $\langle S_z ({\bf r})\rangle$ component of the spin
accumulation can be easily obtained in each probe  and does not
depend on the distance of the collectors.

\

Also in this case we  discuss some devices able to act as spin
filters, here based on the $\beta$-SOC. We carry out a comparison
with the results recently reported in ref.[\onlinecite{hatt}] (the
fundamental parameter in that paper is $\alpha$, corresponding to
$\lambda^2$ in this article, while the ratio $\alpha/l^2$
corresponds to $\lambda^2/l_\omega^2=\omega_{\text{eff}}/\omega$).

In ref.[\onlinecite{hatt}] the variation of $G_{sH}$ was discussed
with the SO coupling strength for $0<\omega_{\text{eff}}/\omega<1$.
The authors showed that $G_{sH}$ increases with
$\omega_{\text{eff}}$, up to the value  $\sim 0.2 \omega$, and then
it tends to saturate. This prediction is in agreement with the
predicted spin polarization in the transverse leads that saturates
between $0.1<\omega_{\text{eff}}<1$, as shown in the right panel of
Fig.(3). Moreover, we predict some oscillations near
$\omega_{\text{eff}} \sim 0.1 \omega$, corresponding to the
quenching oscillations discussed in ref.[\onlinecite{noiq}]

The SOC strengths have been theoretically evaluated for some
semiconductors compounds[\onlinecite{41}]. In a QW patterned in
InGaAs/InP heterostructures,  $\lambda^2$ takes values between $50$
and $150\,{\AA}^2$, hence ($\beta\sim 0.1 \alpha_0$\cite{morozb}).
Thus, one has $\hbar \omega_{\text{eff}}\sim 10^{-6}-10^{-4} eV$,
corresponding to $\omega_{\text{eff}}/\omega\sim 10^{-4}-10^{-3}$.
For GaAs heterostructures, $\lambda^2$ is one order of magnitude
smaller ($\sim 4.4 {\AA}^2$) than in InGaAs/InP, whereas for HgTe
based heterostructures it can be more than three times
larger\cite{HgTe}. However, the lithographical width of a wire
defined in a 2DEG can be as small as $20nm$\cite{kunze}; thus we can
realistically assume that $\omega_{\text{eff}}/\omega$ runs from $1
\times 10^{-6}$ to $1\times 10^{-1}$ (always away from the
saturation region).

 In any case $W$ should be larger than $\lambda_F$, so that
at least one conduction mode is occupied. In this realistic range of
values the spin polarization due to the $\beta$-SOC turns out to be
at least one order of magnitude larger than the one due to the
corresponding value of the $\alpha$-SOC assumed as $\alpha\sim 10
\beta$. Thus we can conclude that, for devices patterned in a 2DEG
at a nanometric scale, the $\beta$ coupling has stronger effects
than the $\alpha$ coupling, as it can be seen by comparing the two
panels of Fig.(3).

\

{\it Conclusions -} In this paper we propose a new scheme of spin
filtering based on nanometric crossjunctions in the presence of
 SO interaction.
The cross geometry seems to be one of the most efficient devices
for the spin filtering, because it acts like an ultra-sensitive
scale, capable of reacting to the smallest variations of both
external and effective fields.
 Here we discussed the SHE in such a small
ballistic device, in which the SOC arises due to the in-plane
confining potential ($\beta$), in contrast to the more widely
studied situation of the Rashba SOC ($\alpha$).

Our approach allows us to make a comparison between the effects
due to each term of SOC, thus we demonstrate that in many cases
the spin filtering based on $\beta$-SOC can be more efficient than
the one based on the $\alpha$-SOC. It happens in nanometric
crossjunction with opportune values of $\lambda$, thus the
feasibility of similar devices  depends on their  size and on the
materials.

\

We acknowledge the support of the grant 2006 PRIN "Sistemi
Quantistici Macroscopici-Aspetti Fondamentali ed Applicazioni di
strutture Josephson Non Convenzionali".


\bibliographystyle{prsty} 

\bibliography{}

\end{document}